# A novel boundary integrated neural networks for in plane fracture mechanics analysis of elastic and piezoelectric materials

Peijun Zhang[1*], Yan Gu[2*], Okyay Altay[1], Chuanzeng Zhang[1, 2*]

[1]*Department of Civil Engineering, University of Siegen, Paul-Bonatz-Str. 9-11, D-57076 Siegen, Germany*

[2]*Faculty of Mechanical Engineering and Mechanics, Ningbo University, Ningbo 315211, PR China*

## Abstract

In this study, we propose a novel approach, termed boundary integrated neural networks (BINNs), for analyzing in-plane crack problems within the framework of linear elastic fracture mechanics. The proposed approach integrates artificial neural networks (ANNs) with classical boundary integral equations (BIEs), enabling an efficient and accurate evaluation of partial differential equations (PDEs) associated with fracture mechanics. Additionally, novel special ANN-based crack-tip elements, the Special crack-tip Neural Networks(SPNNs) are developed to improve the modeling of displacement and stress fields in regions near crack tips. These specialized elements integrate the asymptotic characteristics of fracture mechanics into the neural network framework, ensuring enhanced accuracy in capturing the intricate singularities and steep gradients near the crack tips. Compared to conventional simulation tools in fracture mechanics, the present method offers several distinct advantages. First, by embedding higher-order fracture mechanics principles into the neural networks, the method achieves a more accurate and reliable representation of the near-tip fields, even when using relatively large crack-tip elements. Second, the SPNNs, which incorporates information about varying near-tip singularity orders, improve the method's versatility in solving problems involving complex and diverse crack-tips geometries. Moreover, the method demonstrates excellent computational efficiency due to the dimensionality reduction achieved by employing BIEs.

---

[*] Corresponding Authors: Yan Gu (guyan1913@163.com), Chuanzeng Zhang (c.zhang@uni-siegen.de)

Numerical experiments confirm that the proposed framework serves as a reliable, robust, and accurate tool for addressing fracture mechanics problems, offering substantial advantages over conventional numerical approaches.

***Keywords:*** Artificial Neural Networks; Boundary Integrated Neural Networks; Linear Piezoelectric Fracture Mechanics; Boundary Integral Equations; Coupled Mechanical-Electrical Problems.

**Nomenclature**

| | |
|---|---|
| ANNs | Artificial Neural Networks |
| BEM | Boundary Element Method |
| BINNs | Boundary Integrated Neural Networks |
| BIEs | Boundary Integral Equations |
| COD | Crack Opening Displacement |
| CR | SIFs estimation methods proposed by Cotterell and Rice [1] |
| CRT | SIFs estimation methods, including T-stress terms, proposed by He et al. [2] |
| FEM | Finite Element Method |
| SIFs | Stress Intensity Factors |
| EDIFs | Electric Displacement Intensity Factors |
| N-SIFs | Notch Stress Intensity Factors |
| PINNs | Physics-Informed Neural Networks |
| SPNNs | Special crack-tip Neural Networks |
| PDEs | Partial Differential Equations |
| $K_\mathrm{I}$ | Mode I SIFs |
| $K_\mathrm{II}$ | Mode II SIFs |
| $K_\mathrm{IV}$ | EDIFs |

# 1 Introduction

Fracture mechanics analysis presents a significant challenge within computational mechanics due to the inherent complexities introduced by cracks. The presence of cracks gives rise to singularities in the stress field, where stress values theoretically approach infinity at the crack tips. To address this challenge, various extended or generalized Finite Element Methods (FEM) have been developed. These methods often incorporate specialized crack-tip elements, such as quarter-point elements [3-5], or use enriched elements [6-15] that embed the asymptotic crack-tip field into the finite element formulation. Such approaches have demonstrated high accuracy in linear elastic fracture mechanics analysis.

The Boundary Element Method (BEM) offers an attractive alternative to FEM for fracture mechanics analysis due to its boundary-only discretization and semi-analytical principles. Unlike FEM, which requires volumetric meshing, BEM reduces the dimensionality of the problem, thereby simplifying the modeling process. Over the past decades, researchers have extensively studied the application of BEM to crack problems, exploring both conventional formulations and those enhanced with various enrichment techniques [16-23]. One notable advancement in BEM involves the use of specialized crack-tip elements [24-26]. These elements replace traditional polynomial shape functions with crack-tip-specific shape functions, enabling accurate representation of the local asymptotic behavior near the crack tips. This enhancement significantly improves the accuracy of Stress Intensity Factor (SIF) calculations, a critical parameter in fracture mechanics that governs crack growth and structural integrity.

The introduction of piezoelectric materials into fracture mechanics analysis adds another layer of complexity to the problem, as these materials exhibit coupled mechanical and electrical behaviors, leading to a system of coupled partial differential equations [16, 27-30]. In addition, significant challenges remain in the analysis of piezoelectric fracture mechanics. The process of mesh generation and re-meshing required for simulating crack propagation, particularly in three-dimensional (3D) piezoelectric problems, is still labor-intensive and computationally expensive. Moreover, the handling of complex crack geometries in piezoelectric materials further complicates the analysis, requiring the development of specialized numerical techniques and an increase in computational resources [31]. The accurate prediction of fracture and failure in piezoelectric materials thus requires the continued development of efficient and robust numerical methods capable of dealing with the coupled nature of electrical and mechanical fields in the presence of cracks.

To overcome these challenges, recent advancements in computational mechanics have introduced hybrid methods that combine traditional numerical approaches with machine learning techniques, such as Physics-Informed Neural Networks (PINNs) [32-37] and its variational types [38-46]. PINNs seamlessly integrate information from both physical laws and measurement data by embedding the underlying PDEs and their associated boundary/initial conditions into the loss functions of a neural network. The unknown parameters of the network are optimized by minimizing these loss functions using advanced optimization algorithms. Within the framework of computational fracture mechanics, PINNs have shown great promise in solving complex fracture problems. For instance, Goswami et al. [47] proposed a modified PINNs framework that minimizes the system's variational energy to model brittle fracture using phase-field methods. Similarly, Gu et al. [48, 49] introduced enriched PINNs, which incorporate crack-tip asymptotic functions to improve the accuracy of SIFs predictions in 2D fracture mechanics analysis. Wei et al. [50] developed a physics-informed deep learning model that combines a neural network with the lattice particle method (LPM) to efficiently predict material fracture patterns under arbitrary microstructures and loading conditions. Furthermore, He et al. [51] proposed a PINNs-based framework incorporating sensitive features and life prediction models, enhancing both the training process and the predictive accuracy for multiaxial fatigue life prediction. Chen et al. [52] applied the enhanced PINNs to simulate the crack propagation and predict the fatigue life. By integrating prior knowledge into machine learning frameworks, these methods provide a robust and efficient alternative to conventional computational approaches.

Recently, a novel machine learning framework, known as boundary integrated neural networks (BINNs) [53-56], has been introduced to effectively solve boundary value problems. BINNs combine the capabilities of artificial neural networks with the well-established BEM to efficiently solve PDEs. A distinctive feature of BINNs is its ability to integrate prior knowledge of BIEs into the neural network framework. By leveraging the boundary-only discretization provided by BEM, BINNs reduce the dimensionality of the problem, resulting in enhanced computational efficiency and accuracy. The main idea behind the method can be summarized as follows: First, the original PDEs are reformulated into BIEs by utilizing fundamental solutions or Green's functions, which consist of explicit kernels and unknown boundary quantities. Instead of directly solving the integral equations, a neural network is constructed to approximate these unknown boundary quantities. The network is then trained to satisfy the given BIEs at a set of collocation points. This training process minimizes a loss function, typically derived from the BIEs. Compared to traditional neural network-based methods, BINNs offer several distinct advantages. First, the method seamlessly integrates prior knowledge of BIEs into the training process, focusing solely on satisfying the boundary integral equations. This integration drastically reduces the amount of training data required, leading to a more efficient and faster learning process. Second, by replacing the differential operator in PDEs with an integral operator, BINNs eliminate the need to compute high-order derivatives of the neural networks. This substitution is particularly advantageous, as high-order derivatives are often a source of numerical instability during training. BINNs, therefore, represent a significant advancement over conventional methods by improving both the accuracy and stability of solutions to complex boundary value problems.

In this paper, we propose a novel machine learning framework that integrates BINNs with a specialized ANN-based crack-tip element, the Special crack-tip Neural Networks(SPNNs) to address 2D fracture mechanics problems of piezoelectric materials. The proposed SPNNs are specifically designed to model local stress singularities near crack-tips, enabling accurate representation of the complex displacement and stress fields in near-tip regions. These specialized elements incorporate prior knowledge of fracture mechanics principles, such as those derived from Williams' expansion for crack-tips, into the neural network architecture, ensuring enhanced accuracy in capturing the intricate singularities and steep gradients near the crack tips. The primary objective of this study is to develop a robust, comprehensive, and precise computational framework capable of effectively addressing crack problems across various domain geometries. Through a series of numerical investigations, we aim to demonstrate the effectiveness and versatility of the proposed method, highlighting its ability to handle various crack configurations, including different crack lengths and orientations.

## 2 Problem Statement and Boundary Integral Equations (BIEs)

### 2.1 Problem Statement

Based on the theory of the linear piezoelectricity, the coupled mechanical and electrical equilibrium equations for piezoelectric materials, in the absence of body forces and electrical charges, can be expressed as follows [27, 57, 58]

$$\sigma_{ij,j} = 0, \quad i,j = 1,2, \tag{1}$$

$$D_{i,i} = 0, \quad i = 1,2, \tag{2}$$

with the boundary conditions

$$\begin{cases} u_i = \bar{u}_i & \text{on } \Gamma_u \\ \sigma_{ij} n_j = \bar{t}_i & \text{on } \Gamma_t \end{cases} \quad \text{and} \quad \begin{cases} \phi = \bar{\phi} & \text{on } \Gamma_\phi \\ D_i n_i = -\bar{\omega} & \text{on } \Gamma_\omega \end{cases}, \tag{3}$$

where $D_i$ stands for the electrical displacement vector, $\omega$ and $\phi$ denote the surface charge and the electrical potential, respectively. The aforementioned equations constitute the exact mathematical representation of the elastostatic and piezoelectric fields in 2D solids.

## 2.2 Boundary Integral Equations (BIEs)

By applying the reciprocity theorem, the governing partial differential equations (PDEs) can be transformed into a set of BIEs. The general representation of the BIEs, applicable to both elastostatic and piezoelectric equations, can be expressed as follows

$$\mathbf{C}(P)\mathbf{u}(P) = \int_S \mathbf{U}(P,Q)\mathbf{t}(Q)dS(Q) - \oint_S \mathbf{T}(P,Q)\mathbf{u}(Q)dS(Q), \tag{4}$$

where $P$ and $Q \in \Gamma$ denote the source and observation points, respectively, the jump term $\mathbf{C}(P)$ solely depends on the local geometry at the point $P$ with a specific value of $C_{ij}(P) = \delta_{ij}/2$ for a smooth boundary, $\oint_S$ represents the singular integral with a strong singularity in the sense of the Cauchy principal value (CPV), $\mathbf{u}$ and $\mathbf{t}$ represent the displacement and traction vectors, $\mathbf{U}$ and $\mathbf{T}$ stand for the displacement and traction fundamental solutions, respectively. The explicit expressions of the displacement and traction fundamental solutions for the linear elasticity and linear piezoelectricity can be found in Appendix A. Fundamental solutions

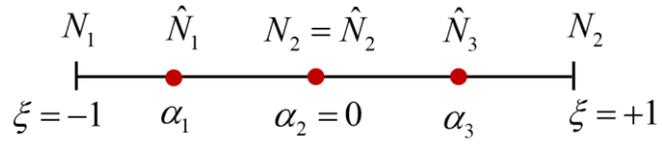

**Fig. 1**. Local coordinate system for discontinuous quadratic elements

Within the BEM framework, the boundary $\Gamma$ of the domain under consideration will be subdivided into a collection of $N$ boundary elements. On these elements, the boundary quantities are approximated by using constant, linear, or quadratic shape functions. When employing the discontinuous quadratic shape functions (see **Fig. 1**), the displacement and traction vectors are expressed as follows:

$$\mathbf{u} = \mathbf{u}^1 \hat{N}_1(\xi) + \mathbf{u}^2 \hat{N}_2(\xi) + \mathbf{u}^3 \hat{N}_3(\xi), \tag{5}$$

$$\mathbf{t} = \mathbf{t}^1 \hat{N}_1(\xi) + \mathbf{t}^2 \hat{N}_2(\xi) + \mathbf{t}^3 \hat{N}_3(\xi), \tag{6}$$

where $\mathbf{u}^k$ and $\mathbf{t}^k$ ($k = 1, 2, 3$) are the physical quantities at the interpolation points, and $\hat{N}_i$ represents the discontinuous interpolation functions.

Based on the aforementioned notations, we can now discretize the BIEs (4) in the following manner:

$$\mathbf{C}(P^m)\mathbf{u}(P^m) = \sum_{n=1}^{N} \sum_{k=1}^{3} \mathbf{t}^{n,k} \int_{S_n} \mathbf{U}(P^m, Q_n(\xi)) N_{kP}(\xi) J(\xi) d\xi \\ - \sum_{n=1}^{N} \sum_{k=1}^{3} \mathbf{u}^{n,k} \oint_{S_n} \mathbf{T}(P^m, Q_n(\xi)) N_{kP}(\xi) J(\xi) d\xi, \tag{7}$$

where $N$ represents the total number of the boundary elements, $P^m$ stands for the $m^{\text{th}}$ collocation/observation point on the boundary, $\mathbf{t}^{n,k}$ and $\mathbf{u}^{n,k}$ are the physical quantities at the $k^{\text{th}}$ interpolation point of the $n^{\text{th}}$ boundary element, $J(\xi)$ is the Jacobian of the transformation from the global coordinates to the local coordinates. By solving the discretized BIEs, one can obtain all the unknown physical quantities on the boundary. Then, the displacement and stress fields in the interior of the considered domain can be evaluated [59]. For further details, interested readers are referred to Refs. [59-65].

The above steps describe the basic procedure of the traditional BEM for solving a boundary value problem. While various BEM-based approaches enjoy the advantages of the easy-meshing and high accuracy, their efficiency in solving large-scale problems becomes a significant challenge. This is primarily due to the fact that the traditional BEM formulation involves dense and non-symmetric system matrices. Although these matrices are smaller in size, they still require a significant amount of memory and computational operations when using direct solvers. Although this bottleneck can be overcome through the development of various acceleration or fast techniques [66-68], it often involves complex mathematical manipulations that are not convenient for researchers and users to implement.

## *2.3 Stress intensity factors and the near-tip displacement and stress fields*

In 2D linear piezoelectric fracture mechanics (LPFM), the electromechanical fields near the crack tip are of central importance. As the crack tip is approached, these fields follow a consistent asymptotic distribution [31, 69]. The mechanical stresses and electric displacements, given by Eq. (8), exhibit a square root singularity, while the electric potential and mechanical displacements, described in Eq. (9), vary with $\sqrt{r}$. The angular functions $f_{ij}(\varphi)$, $g_{ij}(\varphi)$, $d_{ij}(\varphi)$, and $v_{ij}(\varphi)$ depend solely on the material properties [31, 70].

$$\sigma_{ij}(r,\varphi) = \frac{1}{\sqrt{2\pi r}} \left[ K_{\mathrm{I}} f_{ij}^{\mathrm{I}}(\varphi) + K_{\mathrm{II}} f_{ij}^{\mathrm{II}}(\varphi) + K_{\mathrm{IV}} f_{ij}^{\mathrm{IV}}(\varphi) \right]$$
$$D_j(r,\varphi) = \frac{1}{\sqrt{2\pi r}} \left[ K_{\mathrm{I}} g_{ij}^{\mathrm{I}}(\varphi) + K_{\mathrm{II}} g_{ij}^{\mathrm{II}}(\varphi) + K_{\mathrm{IV}} g_{ij}^{\mathrm{IV}}(\varphi) \right]$$
(8)

$$u_i(r,\varphi) = \sqrt{\frac{2r}{\pi}} \left[ K_{\mathrm{I}} d_{ij}^{\mathrm{I}}(\varphi) + K_{\mathrm{II}} d_{ij}^{\mathrm{II}}(\varphi) + K_{\mathrm{IV}} d_{ij}^{\mathrm{IV}}(\varphi) \right]$$
$$\omega(r,\varphi) = \sqrt{\frac{2r}{\pi}} \left[ K_{\mathrm{I}} v_{ij}^{\mathrm{I}}(\varphi) + K_{\mathrm{II}} v_{ij}^{\mathrm{II}}(\varphi) + K_{\mathrm{IV}} v_{ij}^{\mathrm{IV}}(\varphi) \right]$$
(9)

The mechanical stress intensity factors (SIFs) $K_{\mathrm{I}}$ and $K_{\mathrm{II}}$ quantify the intensity of the crack-tip singularity under different loading modes. In piezoelectric materials, an additional parameter, the electric displacement intensity factor (EDIFs) $K_{\mathrm{IV}}$, is introduced to characterize the singularity of the electric displacement field. The K concept is a fundamental framework in LPFM for describing the highly localized crack-tip fields, where the stress follows an inverse square root singularity. The crack tip is typically modeled as a sharp V-notch, with SIFs and EDIFs quantifying the intensity of the crack-tip fields under various loading conditions.

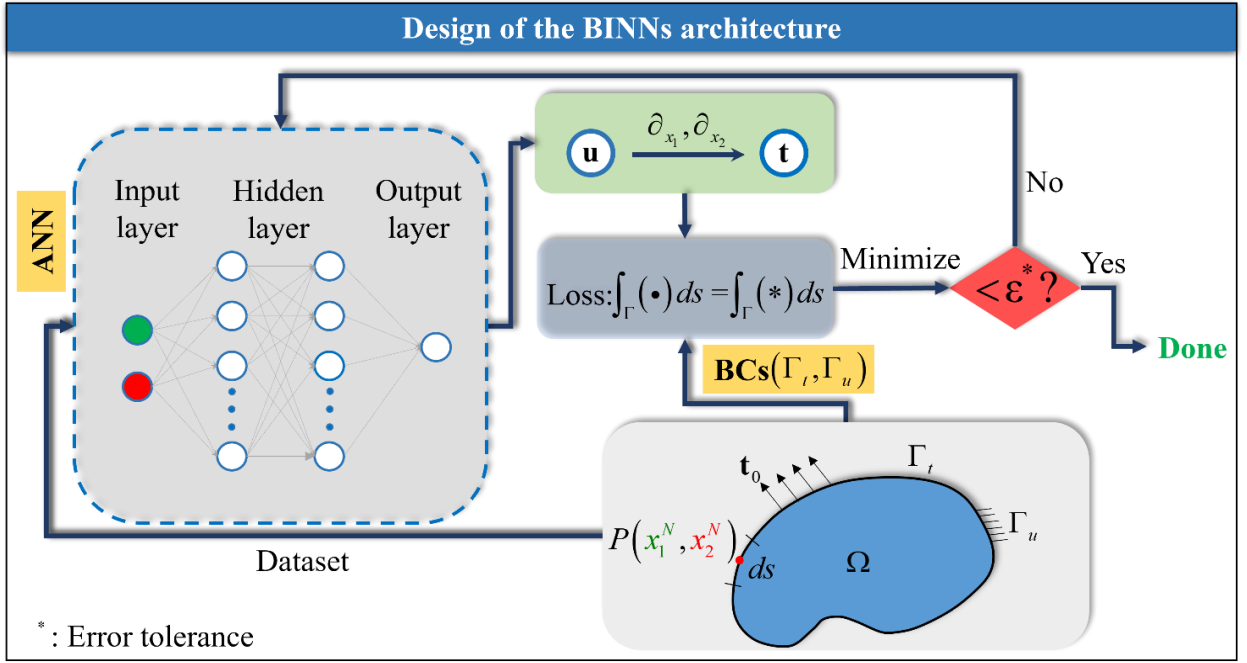

**Fig. 2.** The framework of BINNs for 2D elastostatic problems.

## 3 Artificial Neural Networks and Deep Learning for BIEs

### 3.1 General architecture of BINNs

As a new machine learning framework, BINNs integrate the capabilities of artificial neural networks with the BEM to efficiently solve boundary value problems. Rather than directly tackling the integral equation (13), a neural network is constructed to approximate unknown boundary quantities $\mathbf{u}(Q)$ and $\mathbf{t}(Q)$ in terms of the inputs $x_1$ and $x_2$. This neural network is then trained by satisfying the given BIEs (13) at a set of collocation points. The input to the network is typically a set of boundary points, while the output consists of the predicted boundary values, such as displacement and traction. By leveraging the boundary-only discretization provided by BEM, BINNs reduce the dimensionality of the problem, offering improved computational efficiency.

The framework of BINNs for solving 2D elastostatic problems is illustrated in Fig. 2. In this framework, the target displacement solutions $u_1(x_1, x_2)$ and $u_2(x_1, x_2)$ are approximated using neural networks as follows:

$$u_1(x_1, x_2, \boldsymbol{w}, \boldsymbol{b}) = \sum_{i=1}^{M} w_{1i}^{(1)} \sigma\left(z_i^{(0)}\right) + b_1^{(1)}, \tag{10}$$

$$u_2(x_1, x_2, \boldsymbol{w}, \boldsymbol{b}) = \sum_{i=1}^{M} w_{2i}^{(1)} \sigma\left(z_i^{(0)}\right) + b_2^{(1)}, \tag{11}$$

with

$$z_i^{(0)} = w_{i1}^{(0)} x_1 + w_{i2}^{(0)} x_2 + b_i^{(0)}, \tag{12}$$

Here, $\sigma(\cdot)$ denotes the activation function, $w_{1i}^{(1)}$ and $w_{2i}^{(1)}$ are the weights linking the input unit to the $i$-th neuron unit for $u_1$ and $u_2$ respectively, $b_1^{(1)}$ and $b_2^{(1)}$ are biases associated with the neuron units, and $M$ represents the number of neuron units in the hidden layer. Various activation functions such as Sigmoid, Tanh, Swish, Softplus, Arctan, Mish, and ReLU can be employed based on the problem requirements. To train the trainable parameters $\boldsymbol{w}$ and $\boldsymbol{b}$, the target solutions $u_1(x_1, x_2)$ and $u_2(x_1, x_2)$ are substituted into the integral equations (13), and the following loss function is minimized:

$$L(\boldsymbol{w}, \boldsymbol{b}) = L_{\text{BIE}}^{(1)}(\boldsymbol{w}, \boldsymbol{b}) + L_{\text{BIE}}^{(2)}(\boldsymbol{w}, \boldsymbol{b}), \tag{13}$$

with the notations:

$$L_{\text{BIE}}^{(1)}(\boldsymbol{w}, \boldsymbol{b}) = \frac{1}{N_{total}} \sum_{m=1}^{N_{total}} \left| \frac{u_1(P^m)}{2} - \int_S U_{1j} t_j dS + \oint_S T_{1j} u_j dS \right|^2, \tag{14}$$

$$L_{\text{BIE}}^{(2)}(\boldsymbol{w}, \boldsymbol{b}) = \frac{1}{N_{total}} \sum_{m=1}^{N_{total}} \left| \frac{u_2(P^m)}{2} - \int_S U_{2j} t_j dS + \oint_S T_{2j} u_j dS \right|^2, \tag{15}$$

where $L_{\text{BIE}}^{(1)}(\boldsymbol{w}, \boldsymbol{b})$ and $L_{\text{BIE}}^{(2)}(\boldsymbol{w}, \boldsymbol{b})$ represent two loss terms corresponding to the BIEs formulated for solving 2D elastostatic problems, and $N_{total}$ stands for the total number of collocation/ observation points located along the boundary. In BINNs, the specified boundary conditions are inherently incorporated into the BIEs, meaning that only the unknown boundary quantities need to be approximated. As a result, there is no need to introduce additional weights into the loss function, which are commonly used in traditional machine learning methods to balance gradients across different objective functions. This is another significant advantage of the proposed BINNs approach. There are currently two primary approaches for generating the BINNs numerical framework. The first, introduced by the Sun et al. [55], discretizes the boundary into multiple segments, selecting Gaussian points as the collocation points for training the networks, as shown in equation (16):

$$\mathbf{C}(P^m)\mathbf{u}(P^m) = \sum_{n=1}^{N}\sum_{g=1}^{G}\int_{S_n}\mathbf{U}\left(P^m,Q_n(\xi)\right)\mathbf{t}^{n,g}(\mathbf{x})w^g(\xi)J(\xi)d\xi \\ -\sum_{n=1}^{N}\sum_{g=1}^{G}\oint_{S_n}\mathbf{T}\left(P^m,Q_n(\xi)\right)\mathbf{u}^{n,g}(\mathbf{x})w^g(\xi)J(\xi)d\xi, \quad (16)$$

In contrast, similar to the classical BEM, Zhang et al. [53] adopt quadratic shape functions to represent physical quantities on the boundaries. In this approach, collocation points for training the networks are directly chosen as boundary nodes, as expressed in equation (17):

$$\mathbf{C}(P^m)\mathbf{u}(P^m) = \sum_{n=1}^{N}\sum_{k=1}^{3}\mathbf{t}^{n,k}(\mathbf{x})\int_{S_n}\mathbf{U}\left(P^m,Q_n(\xi)\right)N_{kP}(\xi)J(\xi)d\xi \\ -\sum_{n=1}^{N}\sum_{k=1}^{3}\mathbf{u}^{n,k}(\mathbf{x})\oint_{S_n}\mathbf{T}\left(P^m,Q_n(\xi)\right)N_{kP}(\xi)J(\xi)d\xi, \quad (17)$$

A comparative analysis of the results from both studies reveals no fundamental differences between training on Gaussian points and training directly at boundary nodes. However, training merely on boundary nodes offers several advantages, including smaller datasets, faster training processes, and improved stability.

### 3.2 *Multi-Domain BINNs(M-BINNs)*

Addressing displacement discontinuity across crack surfaces is crucial in fracture mechanics analysis.

The multi-domain BINNs (M-BINNs) method has emerged as a specialized approach to handle this challenge. The core concept of M-BINNs involves dividing the entire region into separate domains, allowing for the explicit treatment of both the upper and lower crack surfaces. By recognizing the distinct behavior of these surfaces, the method effectively captures the displacement jump across the crack. The procedural steps of M-BINNs are straightforward. First, the entire region is divided into distinct domains, each corresponding to either the upper or lower crack surface. Within each domain, local loss functions are defined based on the BIEs. To ensure compatibility between the two subregions, displacement continuity and traction equilibrium conditions along the interface are enforced. These conditions are automatically integrated into the BIEs.

## 4 Novel ANN-based crack-tip elements(SPNNs)

### 4.1 Definitions

Due to the asymptotic behavior of displacements and stresses near the crack tip, traditional quarter-point crack-tip elements are commonly used in finite element and boundary element formulations for modeling cracks in homogeneous elastic materials. These elements accurately capture the $\sqrt{r}$-behavior in the near-tip displacement field, as well as the $1/\sqrt{r}$-variation in the near-tip stress field. However, the accuracy of calculations using quarter-point crack-tip elements is highly dependent on the quality of the mesh. Specifically, refining the mesh around the crack-tip is crucial for obtaining more accurate results. To address this issue, as illustrated in Fig. 3, this paper proposes a novel ANN-based crack-tip element to improve the modeling of displacement and stress fields in regions near crack tips.

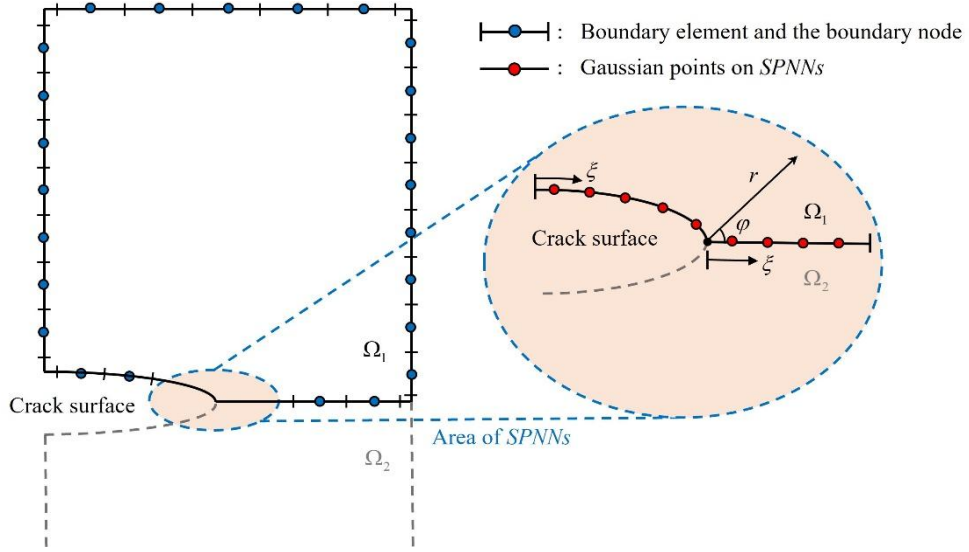

**Fig. 3.** Novel ANN-based crack-tip elements.

Theoretically, the near-tip displacements and tractions must adhere to the following forms:

$$\mathbf{u}(\xi) = \mathbf{d}_1 + \mathbf{d}_2 \cdot r^{1/2} + \mathbf{d}_3 \cdot r^{3/2} + ... \tag{18}$$

$$\mathbf{t}(\xi) = \mathbf{e}_1 + \mathbf{e}_2 \cdot r^{-1/2} + \mathbf{e}_3 \cdot r^{1/2} + ... \tag{19}$$

Where:

$$\mathbf{u}(\xi) = [u_1,\ u_2,\ \phi\ ] \tag{20}$$

$$\mathbf{t}(\xi) = [t_1,\ t_2,\ \omega\ ] \tag{21}$$

By utilizing the geometric relationship between $r$ and the local coordinate $\xi$, we obtain the following relations:

$$\mathbf{u}(\xi) = \overline{\mathbf{d}}_1 + \overline{\mathbf{d}}_2 \cdot (1+\xi)^{1/2} + \overline{\mathbf{d}}_3 \cdot (1+\xi)^{3/2} + ... \tag{22}$$

$$\mathbf{t}(\xi) = \overline{\mathbf{e}}_1 + \overline{\mathbf{e}}_2 \cdot (1+\xi)^{-1/2} + \overline{\mathbf{e}}_3 \cdot (1+\xi)^{1/2} + ... \tag{23}$$

for crack-tip at $\xi = -1$ and

$$\mathbf{u}(\xi) = \overline{\mathbf{d}}_1 + \overline{\mathbf{d}}_2 \cdot (1-\xi)^{1/2} + \overline{\mathbf{d}}_3 \cdot (1-\xi)^{3/2} + ... \tag{24}$$

$$\mathbf{t}(\xi) = \overline{\mathbf{e}}_1 + \overline{\mathbf{e}}_2 \cdot (1-\xi)^{-1/2} + \overline{\mathbf{e}}_3 \cdot (1-\xi)^{1/2} + ... \tag{25}$$

for crack-tip at $\xi=+1$. The SPNNs are referred to as elements, where the displacements and stresses within the elements are characterized using the higher-order asymptotic relations described above. The coefficients $\bar{\mathbf{d}}_i$ and $\bar{\mathbf{e}}_i$ represent the real-valued trainable parameters in these specialized elements, with Gaussian points selected as the training points for the network. In this study, we utilize these specialized elements to model the local displacement and stress fields in the near-tip regions, while for remaining regions, we employ the BINNs, as outlined in Section 3, to model the far-field behavior.

Based on the above analysis, the BIEs for problems involving cracks can now be expressed as follows:

$$\begin{aligned}
&\mathbf{C}(P^i)\mathbf{u}(P^i) \\
&= \left.\begin{aligned}
&\sum_{g=1}^{G}\int_{crack-tip}\mathbf{U}\left(P^i,Q_n(\xi)\right)\left[\bar{\mathbf{e}}_1+\bar{\mathbf{e}}_2\cdot(1\pm\xi)^{-1/2}+...\right]w^g(\xi)J(\xi)d\xi \\
&-\sum_{g=1}^{G}\oint_{crack-tip}\mathbf{T}\left(P^i,Q_n(\xi)\right)\left[\bar{\mathbf{d}}_1+\bar{\mathbf{d}}_2\cdot(1\pm\xi)^{1/2}+...\right]w^g(\xi)J(\xi)d\xi
\end{aligned}\right\} \text{Near-tip-field} \quad (26) \\
&\left.\begin{aligned}
&+\sum_{n=1}^{N}\sum_{k=1}^{3}\mathbf{t}^{n,k}(\mathbf{x})\int_{S_n}\mathbf{U}\left(P^i,Q_n(\xi)\right)N_{kP}(\xi)J(\xi)d\xi \\
&-\sum_{n=1}^{N}\sum_{k=1}^{3}\mathbf{u}^{n,k}(\mathbf{x})\oint_{S_n}\mathbf{T}\left(P^i,Q_n(\xi)\right)N_{kP}(\xi)J(\xi)d\xi,
\end{aligned}\right\} \text{Far-field}
\end{aligned}$$

### 4.2 *Calculation of the singular integrals in SPNNs*

For near-tip-field BIEs, the boundary integrals, as illustrated in Eq. (26), involve the calculation of the terms $\mathbf{u}(\xi)=\bar{\mathbf{d}}_1+\bar{\mathbf{d}}_2\cdot(1\pm\xi)^{1/2}$ and $\mathbf{t}(\xi)=\bar{\mathbf{e}}_1+\bar{\mathbf{e}}_2\cdot(1\pm\xi)^{-1/2}$. The former term is regular and can be accurately calculated using standard Gaussian quadrature, whereas the latter term exhibits singularity of order of $1/\sqrt{1\pm\xi}$ at $\xi=\mp 1$. In this study, a semi-analytical technique based on the integration-by-parts is employed to evaluate the corresponding singular integrals. For demonstration purposes, we focus on the first three terms of the series expansion. Now, let's consider the following integral:

$$\int_{-1}^{1}F(\xi)\mathbf{t}(\xi)d\xi = \int_{-1}^{1}F(\xi)\left[\bar{\mathbf{e}}_1+\bar{\mathbf{e}}_2\cdot(1\pm\xi)^{-1/2}+\bar{\mathbf{e}}_3\cdot(1\pm\xi)^{1/2}\right]d\xi, \qquad (27)$$

where $F(\xi)$ consist of the Jacobian and the displacement fundamental solutions. Now let:

$$\mathbf{T}(\xi) = \overline{\mathbf{e}}_1 \cdot \xi \pm 2 \cdot \overline{\mathbf{e}}_2 \cdot (1 \pm \xi)^{1/2} \pm 2/3 \cdot \overline{\mathbf{e}}_3 \cdot (1 \pm \xi)^{3/2}, \tag{28}$$

and note the following relationship:

$$\mathbf{T}'(\xi) = \mathbf{t}(\xi). \tag{29}$$

The integral (36) can be evaluated as:

$$\int_{-1}^{1} F(\xi) \mathbf{t}(\xi) d\xi = \left[\mathbf{T}(\xi) F(\xi)\right]_{-1}^{1} - \int_{-1}^{1} \mathbf{T}(\xi) F'(\xi) d\xi, \tag{30}$$

where the second integral on the right-hand side is now regular and can be accurately evaluated using the standard Gaussian quadrature. It should be noted that when the collocation point $P$ lies within the integration element, the function $F(\xi)$, which contains the displacement fundamental solutions, exhibits a weak-singularity of order $O(lnr)$. Furthermore, its derivative $F'(\xi)$ demonstrates a strong-singularity of order $O(1/r)$. To handle integrals with strong singularities, various algorithms have been developed and extensively applied within the BEM community. In this study, we adopt the method proposed by Guiggiani and Casalini [60] for accurately computing such singular integrals.

### 4.3 Extraction of Stress Intensity Factors and Electric Displacement Intensity Factors

In the case where the crack surface lies at $\theta = \pi$, the SIFs and EDIFs and can be computed via:

$$K_{\mathrm{I}} = \lim_{\substack{r \to 0 \\ \theta = 0}} \sqrt{2\pi r} \{t_2(r)\}, \quad K_{\mathrm{II}} = \lim_{\substack{r \to 0 \\ \theta = 0}} \sqrt{2\pi r} \{t_1(r)\}, \quad K_{\mathrm{IV}} = \lim_{\substack{r \to 0 \\ \theta = 0}} \sqrt{2\pi r} \{\omega(r)\} \tag{31}$$

Due to the use of the specialized crack-tip elements, which theoretically incorporate the asymptotic displacement and stress fields near the crack-tip, the SIFs and EDIFs in this study can be calculated directly at points very close to the crack tip.

## 5 Numerical Investigations

This section presents the numerical analysis of five benchmark examples to evaluate the performance of BINNs combined with the SPNNs. The first three examples assess the algorithm's accuracy in elastic fracture mechanics by comparing the results with reference solutions. The first example also includes a sensitivity analysis of the BINNs architecture and the effect of crack-tip element size. The fourth and fifth examples extend the analysis to piezoelectric materials.

In all benchmark examples presented in this section, the total number of training points is relatively small, typically on the order of a few hundred. All numerical experiments are conducted using the Python package *TensorFlow* 2.0, along with the limited-memory BFGS optimization algorithm. The computations are performed on a system equipped with an Intel i7 2.90 GHz CPU, 32 GB of RAM, a 500 GB hard drive, and running the Windows 11 operating system. The resulting relative error is computed using the formula: $|K_{num} - K_{exact}|/|K_{exact}|$, where $K_{num}$ and $K_{exact}$ denote the numerical and exact SIF solutions, respectively.

### 5.1 *An edge-cracked plate under a remote tensile stress loading*

The first example investigates the performance of the proposed framework in solving a pure mode I crack problem under plane strain conditions. We perform a fracture mechanics analysis of an edge-cracked plate under a tensile stress loading, as depicted in Fig. 4 (a). The boundaries and far-field interfaces are discretized into 180 discontinuous quadratic elements, where the boundary solutions **u** and **t** are trained using BINNs.

#### 5.1.1 *Sensitivity analysis of BINNs architecture*

Firstly, we delve into the sensitivity of our numerical results concerning on the set of the neural networks in the far field, where the conventional BINNs are implemented. Three crucial factors are hereby considered: the type of applied activation functions, the number of hidden layers and the neurons utilized in each hidden layer. Here we vary the number of neurons per hidden layer within the range of 4 to 32 and the number of hidden layers from 1 to 4. The selected activation functions for investigation include "Tanh", "Sigmoid", "Relu" and "Swish". Here, we focus on comparing the normalized Crack Opening Displacement (COD) at the crack center between the present method and the analytical solution given by [69, 71, 72]

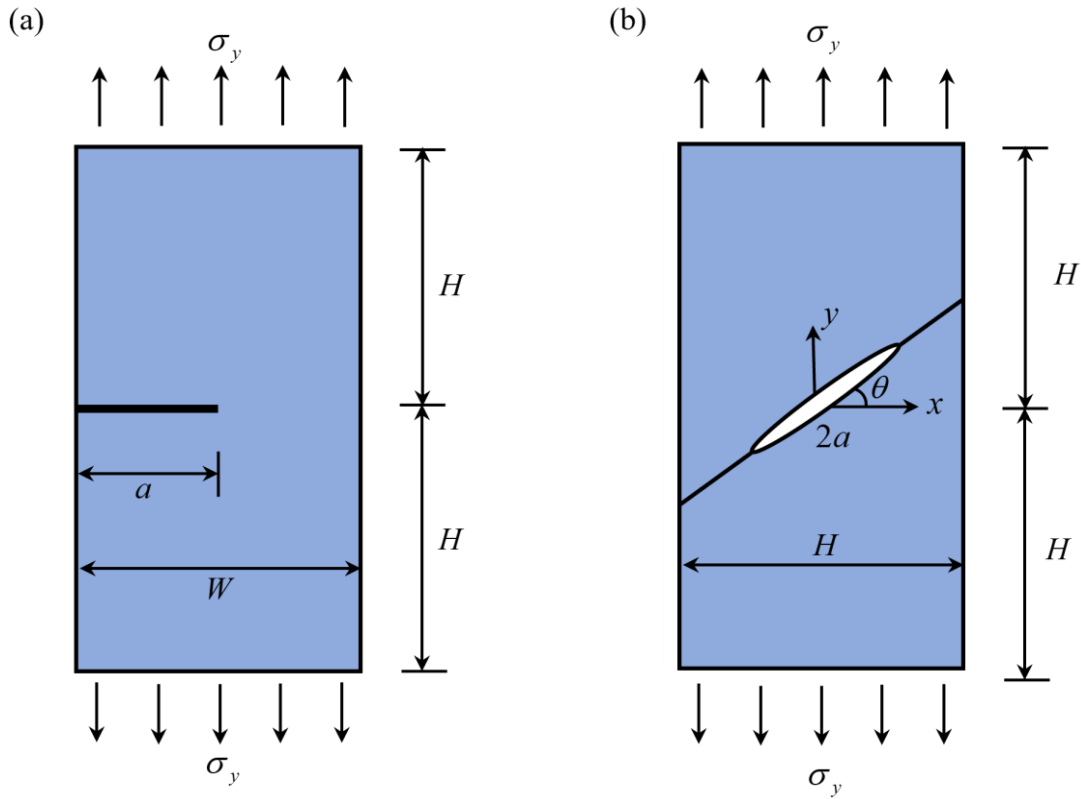

**Fig. 4**. (a). An edge cracked finite plate subjected to tensile stress loading. (b). A finite plate with a slant crack under a tensile stress loading

In Fig. 5, the analysis reveals that the accuracy of computed COD remains consistently stable for the "Relu", "Tanh" and "Swish" activation functions, even when applied to shallow neural networks. Conversely, the "Sigmoid" activation function yields less accurate outcomes. Notably, among the considered activation functions, "Tanh" emerges as a standout performer, displaying comparatively superior accuracy and stability. It is important to note that this sensitivity analysis is tailored to a straightforward configuration. For scenarios involving complex geometry or mixed crack opening modes, a new sensitivity analysis becomes imperative to ensure robust conclusions. In the subsequent examples featuring simple configurations, the model employs three hidden layers with 20 neurons each for consistency in experimentation.

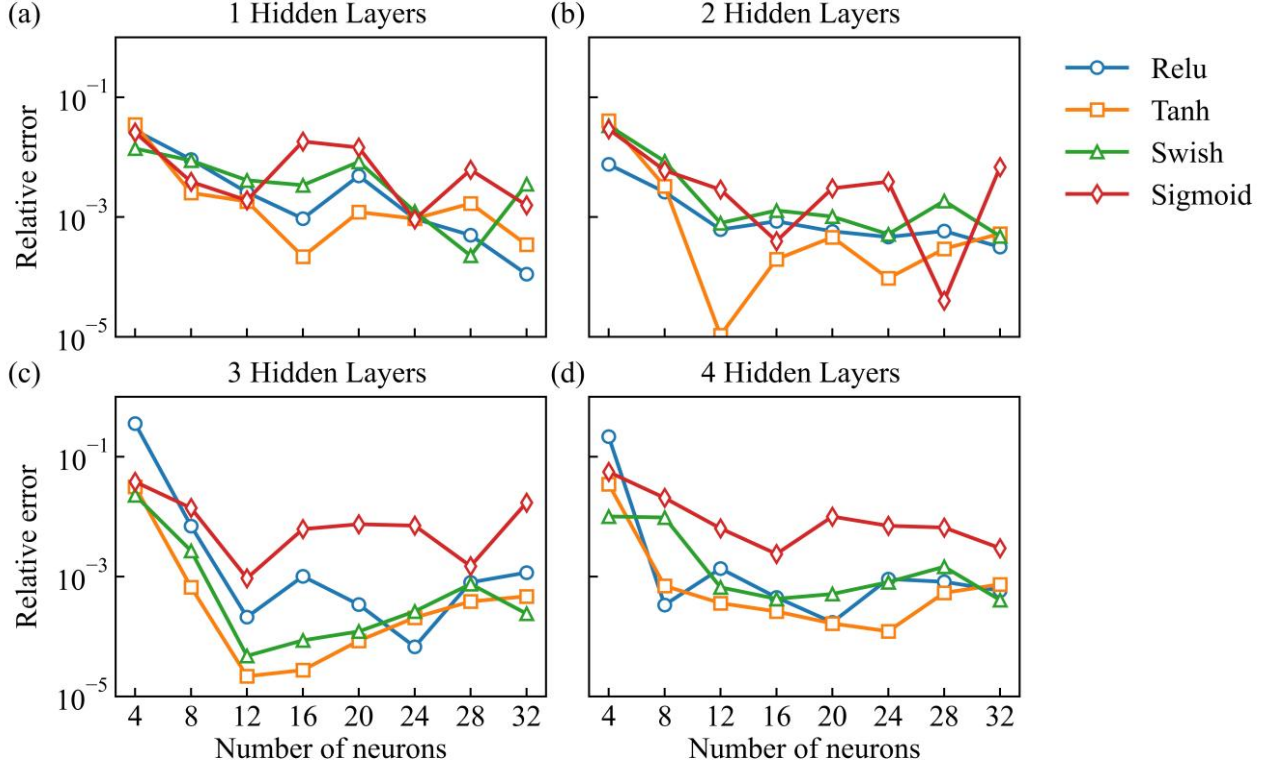

**Fig. 5.** Variation of the relative errors with respect to the number of neurons in each hidden layer, the applied activation functions and with: (a)1 hidden layers, (b) 2 hidden layers, (c) 3 hidden layers, (d) 4 hidden layers.

*5.1.2   Sensitivity analysis of SPNNs size*

In this subsection, we specifically investigate the impact of varying the size of SPNNs within a range of 10% to 80% of the crack length, as illustrated exemplarily in Fig. 6. The boundaries and far-field interfaces are discretized into 80 discontinuous quadratic elements. The primary focus is to assess the SPNN's ability to accurately represent both displacement and stress fields, extending from the crack-tip region to the far field, while maintaining accuracy in predicting SIFs. We systematically vary the crack-length ratio from 0.2 to 0.5 and consider higher order terms in SPNNs ranging from 2 to 7.

Additionally, we conduct a comparative analysis with the BEM utilizing the special crack-tip element proposed by Gu [24]. In this comparison, we investigate whether maintaining the same size for the special crack-tip element as that of SPNNs could still results in sustained accuracy in predicting SIFs. The empirical solution (accuracy < 0.5%) of the mode I SIF $K_\mathrm{I}$ in this problem is given in [73].

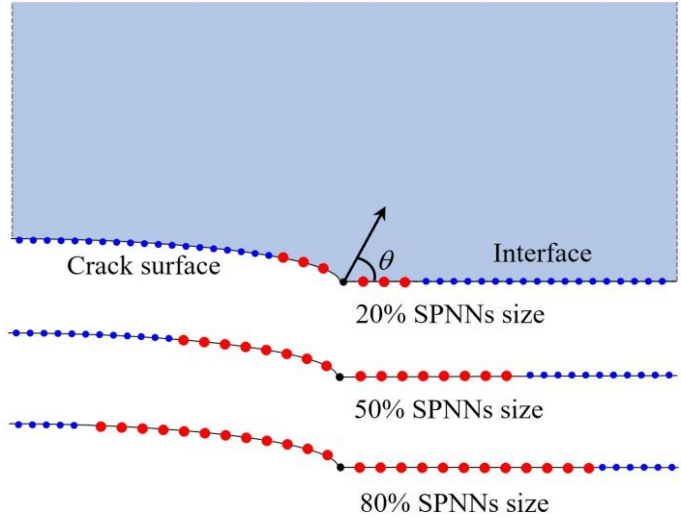

**Fig. 6.** Implementation of varying SPNNs sizes

As evident from Fig. 7, the accuracy of computed SIFs using two-term crack-tip element is notably inferior compared to models incorporating higher order terms. This discrepancy arises because the single asymptotic terms $\mathbf{d}_2$ and $\mathbf{e}_2$ are valid primarily as $r$ tends to zero, insufficient to describe displacement and stress fields at finite distances from the crack-tip. This limitation becomes more pronounced as the SPNNs size increases, indicating that the square root behavior of displacement and the inverse square root singularity of stress are only reliable in the immediate vicinity of the crack-tip.

Fig. 7 also highlights the necessity of incorporating more than three higher-order terms in crack-tip element as their size increases, ensuring an accurate representation of displacement and stress fields over finite distances from the crack tip. Even when the SPNNs size extends to 80% of the crack length, higher-order terms continue to enhance accuracy. In contrast, the accuracy of computed SIFs in the BEM framework [24] deteriorates as the size of the crack-tip element grows, particularly for longer crack lengths. Unlike BEM, SPNNs with more than three higher-order terms remain unaffected by element size, eliminating the need for mesh refinement near the crack tip.

Nevertheless, increasing the number of higher-order terms introduces more trainable parameters, which extends training time and creates a trade-off between accuracy and computational efficiency. As complexity grows, this balancing issue becomes more pronounced, potentially reducing overall efficiency.

In practical engineering applications, the typical error margin in SIFs calculations ranges from 1% to 3%. Within this range, using four or five higher-order terms is generally sufficient to achieve accurate and reliable results without incurring excessive computational costs. Therefore, in the following numerical examples, we fix the number of higher-order terms at five and use 50% of the SPNNs size.

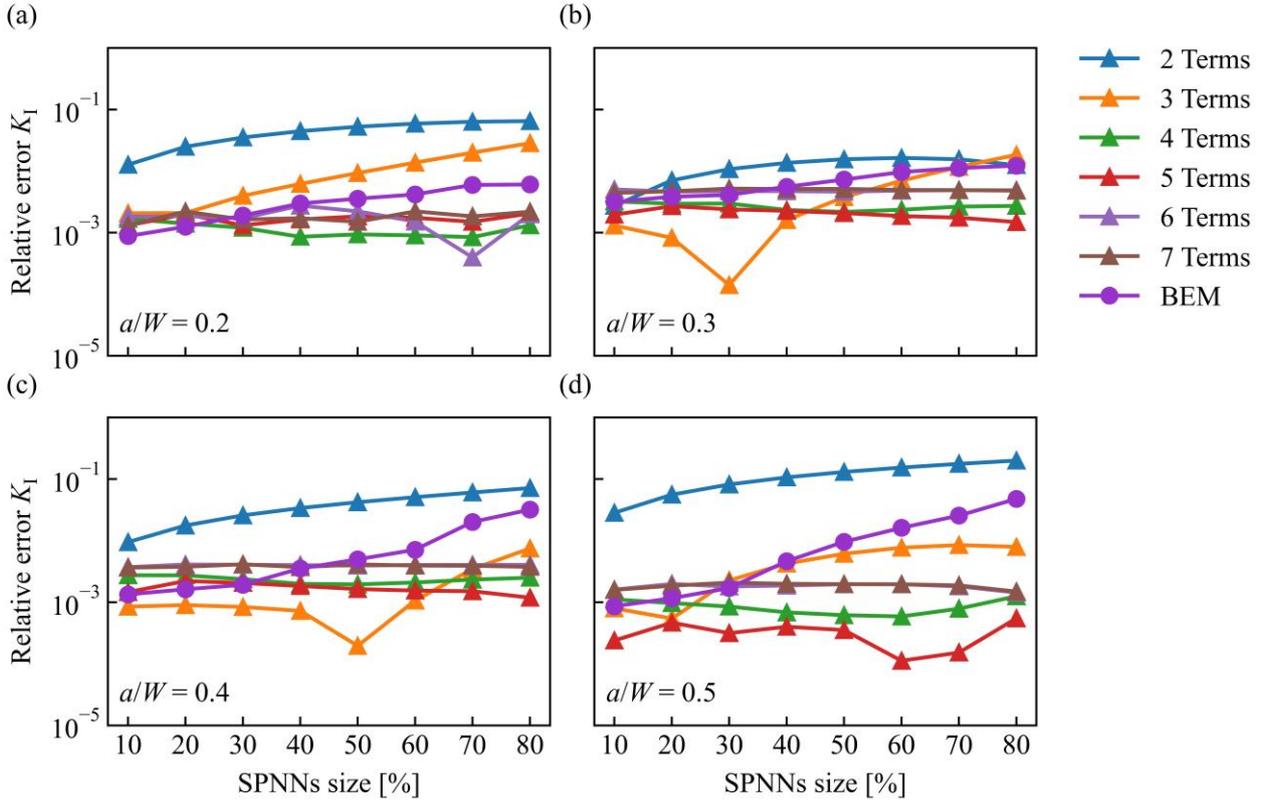

**Fig. 7.** Relative errors of $K_\mathrm{I}$ in an edge-cracked plate under a remote tensile stress loading with varying SPNNs sizes (BINNs with quadratic elements). The crack-length ratios are taken as: (a) $a/W = 0.2$, (b) $a/W = 0.3$, (c) $a/W = 0.4$, (d) $a/W = 0.5$.

### 5.1.3 Stress intensity factor solutions

Fig. 8 illustrates the variation of the estimated mode I SIF $K_\mathrm{I}$ with respect to the relative plate width $a/W$. The results demonstrate that the SIF values calculated using the present method are in excellent agreement with the corresponding empirical results across a wide range of relative plate widths.

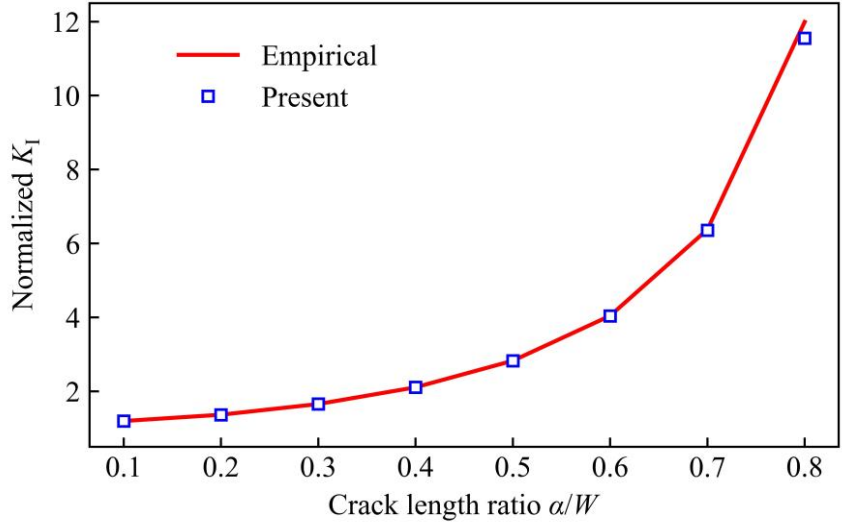

**Fig. 8.** Variation of the estimated normalized $K_I$ with the varying crack length ratio $a/W$.

## 5.2 A finite plate with a slant crack

This example demonstrates the effectiveness of the proposed framework in analyzing mixed mode crack problems, specifically focusing on SIFs on a finite plate with a central slant crack. The crack-length ratio $2a/H$ ranges from 0.1 to 0.8, with crack slant angles $\theta$ set at 15°, 30°, 45°, and 60°. The problem is assumed to be in 2D plane strain condition with a Poisson's ratio $v$ of 0.3.

The network architecture comprises 3 fully-connected hidden layers, each with 30 neurons, trained over 20000 iterations. Mode I and mode II SIFs ($K_I$ and $K_{II}$) are computed for different $\theta$ and $2a/H$ values and compared with numerical results obtained from the BEM by Gu [24] and Kitagawa [74].

**Table 1.** Normalized $K_I$ for slant cracked plate under tensile stress loading.

| $\theta$ | Method | $2a/H$ | | | | | | | |
|---|---|---|---|---|---|---|---|---|---|
| | | 0.1 | 0.2 | 0.3 | 0.4 | 0.5 | 0.6 | 0.7 | 0.8 |
| 15° | Present | 0.9384 | 0.9575 | 0.9898 | 1.0394 | 1.1119 | 1.2174 | 1.3770 | 1.6475 |
| | Gu[72] | 0.9380 | 0.9573 | 0.9899 | 1.0397 | 1.1122 | 1.2178 | 1.3774 | 1.6414 |
| | Kitagawa[74] | 0.9391 | 0.9577 | 0.9904 | 1.0402 | 1.1128 | 1.2183 | 1.3780 | 1.6530 |
| 30° | Present | 0.7570 | 0.7729 | 0.8021 | 0.8453 | 0.9044 | 0.9830 | 1.0897 | 1.2426 |
| | Gu[72] | 0.7548 | 0.7728 | 0.8023 | 0.8454 | 0.9043 | 0.9836 | 1.0909 | 1.2433 |

| | Kitagawa[74] | 0.7557 | 0.7730 | 0.8025 | 0.8456 | 0.9046 | 0.9840 | 1.0910 | 1.2450 |
| 45° | Present | 0.5059 | 0.5186 | 0.5413 | 0.5723 | 0.6119 | 0.6607 | 0.7199 | 0.7927 |
| | Gu[72] | 0.5038 | 0.5181 | 0.5406 | 0.5718 | 0.6118 | 0.6611 | 0.7213 | 0.7949 |
| | Kitagawa[74] | 0.5046 | 0.5181 | 0.5406 | 0.5719 | 0.6119 | 0.6611 | 0.7210 | 0.7950 |
| 60° | Present | 0.2534 | 0.2623 | 0.2748 | 0.2916 | 0.3115 | 0.3350 | 0.3610 | 0.3887 |
| | Gu[72] | 0.2522 | 0.2605 | 0.2734 | 0.2900 | 0.3105 | 0.3332 | 0.3591 | 0.3891 |
| | Kitagawa[74] | 0.2527 | 0.2605 | 0.2730 | 0.2896 | 0.3099 | 0.3332 | 0.3590 | 0.3880 |

**Table 2**. Normalized $K_{\mathrm{II}}$ for an edge slant cracked plate under remote normal stress loading.

| $\theta$ | Method | $2a/H$ | | | | | | | |
| --- | --- | --- | --- | --- | --- | --- | --- | --- | --- |
| | | 0.1 | 0.2 | 0.3 | 0.4 | 0.5 | 0.6 | 0.7 | 0.8 |
| 15° | Present | 0.2499 | 0.2506 | 0.2519 | 0.2554 | 0.2615 | 0.2722 | 0.2930 | 0.3314 |
| | Gu[72] | 0.2458 | 0.2503 | 0.2522 | 0.2553 | 0.2611 | 0.2719 | 0.2923 | 0.3325 |
| | Kitagawa[74] | 0.2502 | 0.2510 | 0.2527 | 0.2560 | 0.2619 | 0.2725 | 0.2900 | 0.3070 |
| 30° | Present | 0.4334 | 0.4350 | 0.4416 | 0.4491 | 0.4612 | 0.4800 | 0.5080 | 0.5529 |
| | Gu[72] | 0.4171 | 0.4345 | 0.4386 | 0.4436 | 0.4608 | 0.4715 | 0.5084 | 0.5520 |
| | Kitagawa[74] | 0.4339 | 0.4367 | 0.4417 | 0.4497 | 0.4617 | 0.4800 | 0.5080 | 0.5500 |
| 45° | Present | 0.5025 | 0.5065 | 0.5160 | 0.5286 | 0.5455 | 0.5669 | 0.5940 | 0.6289 |
| | Gu[72] | 0.5039 | 0.5072 | 0.5186 | 0.5287 | 0.5448 | 0.5672 | 0.5941 | 0.6290 |
| | Kitagawa[74] | 0.5018 | 0.5072 | 0.5162 | 0.5290 | 0.5458 | 0.5674 | 0.5950 | 0.6300 |
| 60° | Present | 0.4346 | 0.4414 | 0.4510 | 0.4653 | 0.4825 | 0.5014 | 0.5230 | 0.5475 |
| | Gu[72] | 0.4337 | 0.4416 | 0.4511 | 0.4658 | 0.4838 | 0.5015 | 0.5239 | 0.5481 |
| | Kitagawa[74] | 0.4352 | 0.4417 | 0.4521 | 0.4660 | 0.4827 | 0.5022 | 0.5240 | 0.5490 |

### *5.3 A kinked edge crack in a semi-infinite plate subjected to remote tensile stress loading*

The problem configuration, illustrated in Fig. 9, involves a kink length $l$ fixed at a ratio of $l/a_0 = 0.1$, with kink angles $\theta$ varying from -60° to 60° in 10° increments. This study compares the normalized SIFs obtained using the present method with those from reference methods, namely CR [1], CRT [2] and finite element solutions (FEM).

Fig. 10 presents a comprehensive comparison of the SIF values for $\gamma = 0°$, $\gamma = 30°$ and $\gamma = 60°$ estimated by various methods. The solutions obtained using the present method demonstrate excellent agreement with the finite element solutions, consistently providing more accurate results compared to CR[1] and CRT[2] methods across the entire range of kink angles considered. The findings underscore the potential of the present method as a powerful tool for analyzing nonlinear cracks in semi-infinite plates.

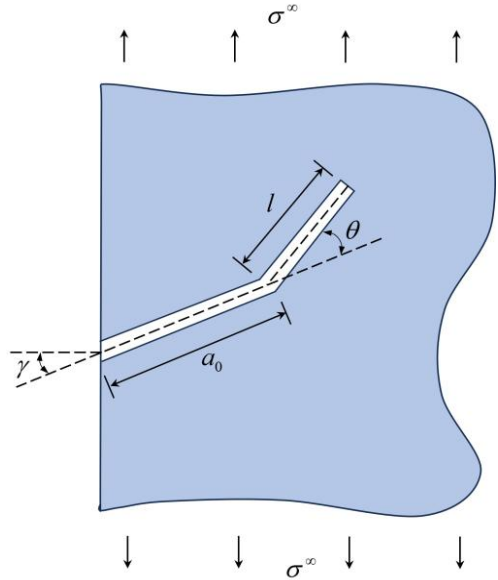

**Fig. 9.** An kinked edge crack in a semi-infinite domain subjected to tensile stress loading

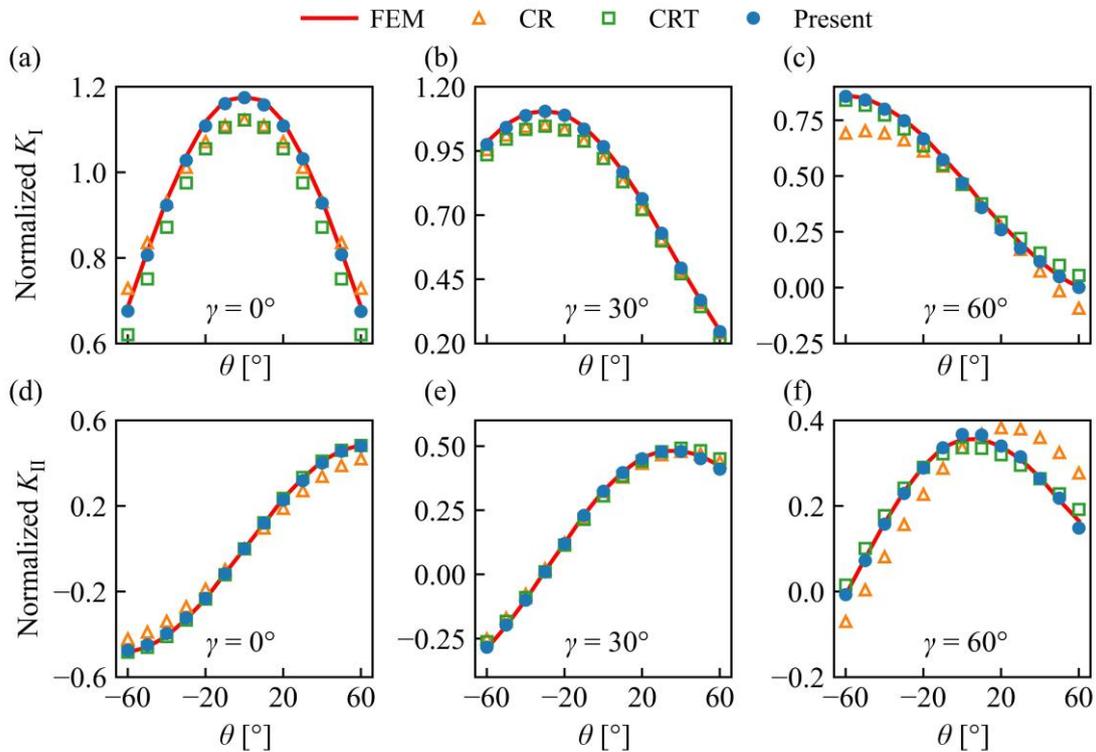

**Fig. 10.** Comparison of solutions for an edge kinked crack in a semi-infinite domain under remote tension

with varying kink angles $\theta$ using present method, FEM, CR and CRT: (a) normalized $K_\mathrm{I}$, $\gamma = 0°$, (b) normalized $K_\mathrm{I}$, $\gamma = 30°$, (c) normalized $K_\mathrm{I}$, $\gamma = 60°$, (d) normalized $K_\mathrm{II}$, $\gamma = 0°$, (e) normalized $K_\mathrm{II}$, $\gamma = 30°$, (f) normalized $K_\mathrm{II}$, $\gamma = 60°$

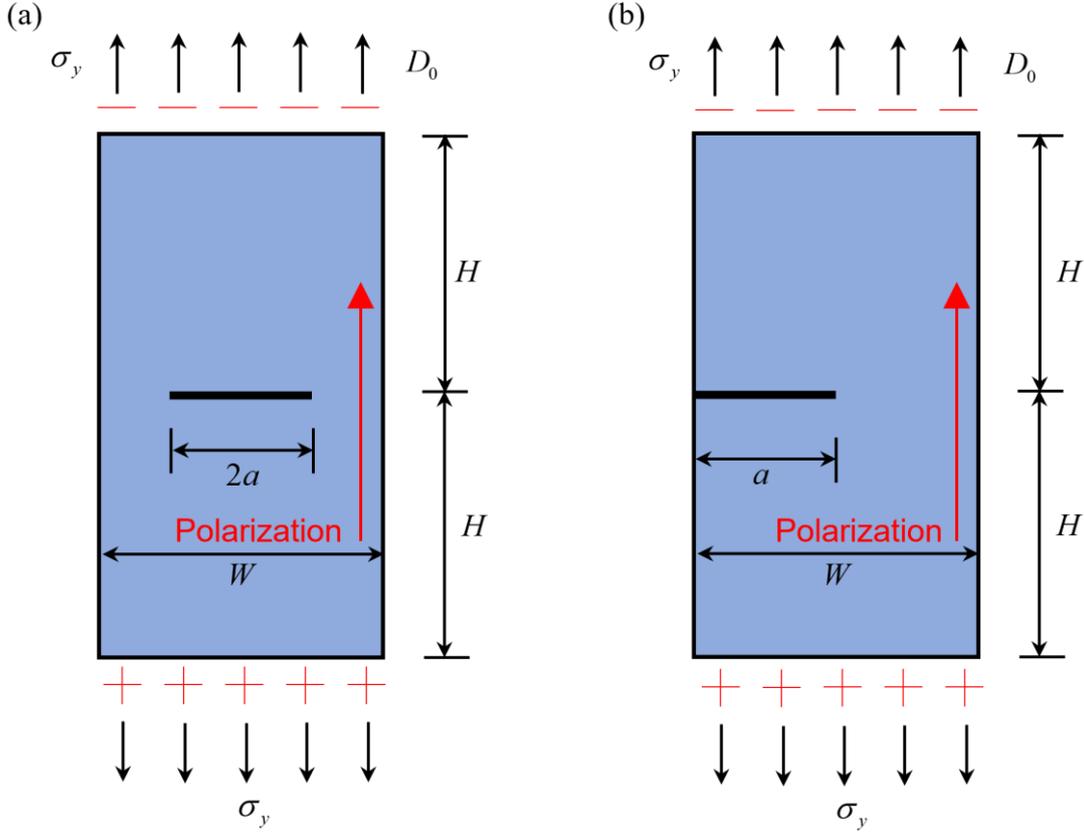

**Fig. 11.** (a). A central cracked piezoelectric strip (b). An edge cracked piezoelectric strip under a tensile stress loading and electric displacement

### *5.4 A center-cracked piezoelectric strip under tensile stress loading and electric displacement*

In the following analysis, we examine a center-cracked PZT-5H piezoelectric strip subjected to both tensile stress and electric displacement loading. The values of the material constants are given by $c_{11} = 12.6$, $c_{13} = 8.41$, $c_{33} = 11.7$, $c_{44} = 2.3$, $e_{31} = -6.5$, $e_{15} = 17.44$, $\epsilon_{11} = 15.03$, $\epsilon_{33} = 13.0$, where the elastic constants $c_{ij}$ are in $10^{10}$ N/m$^2$, the piezoelectric constants $e_{ij}$ in $10^{10}$ C/m$^2$, and the permittivity $\epsilon_{ij}$ in $10^{-9}$ F/m. Our numerical study focuses on the region where electric displacement loads are applied, considering specific values of $D_0 / (e_{33}/c_{33}) \sigma_y = 0, 1, 2, 3$. These values are chosen to encompass both negative and positive electric fields, allowing for a thorough investigation of their influence on crack behavior.

Additionally, based on the findings of Wang and Mai [75], which suggest that impermeable crack boundary conditions offer a more realistic representation for engineering applications, we adopt these conditions in our analysis. This ensures alignment with established theoretical principles and enhances the applicability of our results. Consequently, all subsequent simulations and discussions are conducted under the assumption of impermeable crack boundary conditions.

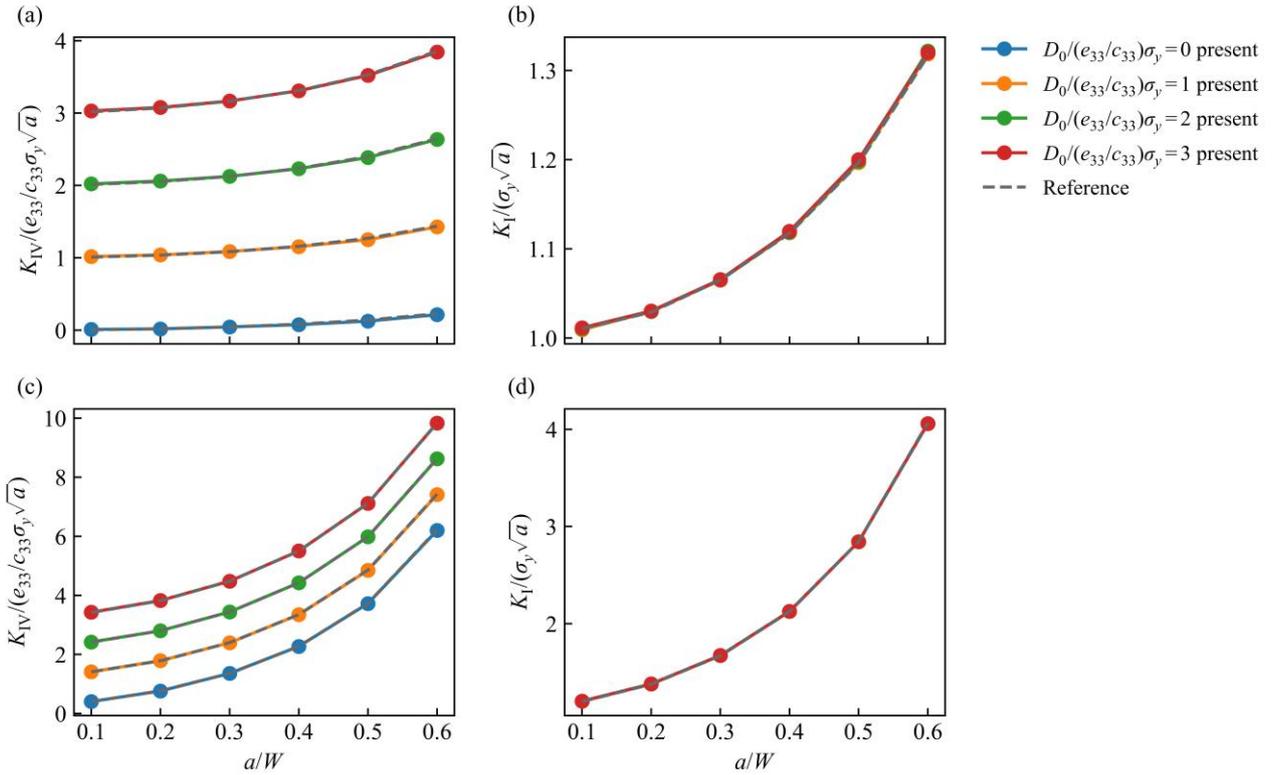

**Fig. 12.** An kinked edge crack in a semi-infinite domain subjected to tensile stress loading

The results illustrated in Fig. 12 (a) and (b) show the distribution of the electric displacement intensity factor $K_{IV}$ and stress intensity factor $K_I$ for a central-cracked piezoelectric strip under uniform tensile and electric displacement loads. The results calculated using the present BINNs framework closely match the reference solutions from [76], demonstrating the accuracy of this approach for analyzing cracked piezoelectric materials.

Fig. 12 (a) illustrates the variation of the electric displacement intensity factor with crack length. For an impermeable crack surface, the electric displacement field at the crack tip increases with the applied electric displacement load. Notably, even in the absence of external electric displacement loading, the electric displacement intensity factor $K_{IV}$ remains non-zero. This demonstrates that purely mechanical loading can induce electric field concentration at the crack tip in the finite plate. In contrast, mechanical and electric displacement intensity factors are entirely uncoupled in infinite piezoelectric media, where the electric displacement intensity factor remains zero under purely mechanical loading [28]. Fig. 12 (b) highlights that the stress intensity factors exhibit negligible sensitivity to the applied electric displacement load.

*5.5 An edge-cracked piezoelectric strip under tensile stress loading and electric displacement*

In the final example, an edge-cracked piezoelectric strip subjected to tensile stress and electric displacement loading is considered. The material properties are identical to those in the previous example. Four loading scenarios are considered, $D_0 / (e_{33} / c_{33}) \sigma_y = 0, 1, 2, 3$, including both positive and negative electric fields.

The computed EDIFs and SIFs are compared with reference solutions [76], as shown in Fig. 12 (c) and (d). The results confirm the accuracy and robustness of the proposed approach for analyzing cracked piezoelectric materials.

# 6 Conclusions

In this paper, we have presented a novel machine learning platform that combines BINNs and SPNNs to address two-dimensional linear piezoelectric fracture mechanics problems. Through a series of numerical investigations, we have demonstrated the effectiveness, versatility, and robustness of the proposed method in handling a wide range of crack configurations across various domains.

The seamless integration of BINNs and SPNNs has enabled accurate representation of complex stress fields, and the accuracy of the platform has been validated against well-established analytical solutions and high-fidelity numerical methods. The method has shown potential for solving complex, real-world fracture mechanics challenges, offering advantages over traditional computational methods in terms of efficiency and adaptability.

The modular structure of the framework allows for easy extension and customization, opening up opportunities for future research and development in computational fracture mechanics. In conclusion, the proposed machine learning platform represents a powerful and versatile tool for addressing linear elastic fracture mechanics problems, making it a valuable asset for researchers and practitioners in the field.

## Acknowledgements

The work described in this paper was supported by the National Natural Science Foundation of China (Nos. 11872220, 12111530006), and the Natural Science Foundation of Shandong Province of China (Nos. 2019KJI009, ZR2021JQ02).

## Data Availability Statement

The data that support the findings of this study are available from the corresponding author upon reasonable request.

## Appendix A. Fundamental solutions

For 2D elastostatic problems, the fundamental solutions are given by

$$U_{ij}(P,Q) = \frac{1}{8\pi\mu(1-\nu)}\left[(4\nu-3)\ln(r)\delta_{ij} + r_{,i}r_{,j}\right], \tag{A32}$$

$$T_{ij}(P,Q) = -\frac{1}{4\pi(1-\nu)r}\left\{\frac{\partial r}{\partial n}\left[(1-2\nu)\delta_{ij} + 2r_{,i}r_{,j}\right] - (1-2\nu)(r_{,i}n_j - r_{,j}n_i)\right\}, \tag{A33}$$

where $\mu$ represents the shear modulus while $\nu$ denotes the Poisson's ratio, $\mathbf{n} = (n_1, n_2)$ represents the unit outward normal vector at the point $Q$, $r = |P-Q|$ stands for the Euclidean distance between the points $P$ and $Q$. For further details, interested readers are referred to Refs. [65, 77, 78].

For 2D piezoelectric problems, the fundamental solutions are given by

$$T_{11} = (k_{11}^1 A_1 \frac{\hat{x}}{r_1^2} + k_{11}^2 A_2 \frac{\hat{x}}{r_2^2} + k_{11}^3 A_3 \frac{\hat{x}}{r_3^2})n_1 + (k_{14}^1 A_1 \frac{\hat{y}_1}{r_1^2} + k_{14}^2 A_2 \frac{\hat{y}_2}{r_2^2} + k_{14}^3 A_3 \frac{\hat{y}_3}{r_3^2})n_2, \tag{A34}$$

$$T_{12} = (k_{14}^1 A_1 \frac{\hat{y}_1}{r_1^2} + k_{14}^2 A_2 \frac{\hat{y}_2}{r_2^2} + k_{14}^3 A_3 \frac{\hat{y}_3}{r_3^2})n_1 + (k_{12}^1 A_1 \frac{\hat{x}}{r_1^2} + k_{12}^2 A_2 \frac{\hat{x}}{r_2^2} + k_{12}^3 A_3 \frac{\hat{x}}{r_3^2})n_2, \tag{A35}$$

$$T_{13} = -(k_{15}^1 A_1 \frac{\hat{y}_1}{r_1^2} + k_{15}^2 A_2 \frac{\hat{y}_2}{r_2^2} + k_{15}^3 A_3 \frac{\hat{y}_3}{r_3^2})n_1 - (k_{13}^1 A_1 \frac{\hat{x}}{r_1^2} + k_{13}^2 A_2 \frac{\hat{x}}{r_2^2} + k_{13}^3 A_3 \frac{\hat{x}}{r_3^2})n_2, \tag{A36}$$

$$T_{21} = (k_{21}^1 B_1 \frac{\hat{y}_1}{r_1^2} + k_{21}^2 B_2 \frac{\hat{y}_2}{r_2^2} + k_{21}^3 B_3 \frac{\hat{y}_3}{r_3^2})n_1 + (k_{24}^1 B_1 \frac{\hat{x}}{r_1^2} + k_{24}^2 B_2 \frac{\hat{x}}{r_2^2} + k_{24}^3 B_3 \frac{\hat{x}}{r_3^2})n_2, \tag{A37}$$

$$T_{22} = (k_{24}^1 B_1 \frac{\hat{x}}{r_1^2} + k_{24}^2 B_2 \frac{\hat{x}}{r_2^2} + k_{24}^3 B_3 \frac{\hat{x}}{r_3^2})n_1 + (k_{22}^1 B_1 \frac{\hat{y}_1}{r_1^2} + k_{22}^2 B_2 \frac{\hat{y}_2}{r_2^2} + k_{22}^3 B_3 \frac{\hat{y}_3}{r_3^2})n_2, \tag{A38}$$

$$T_{23} = -(k_{25}^1 B_1 \frac{\hat{x}}{r_1^2} + k_{25}^2 B_2 \frac{\hat{x}}{r_2^2} + k_{25}^3 B_3 \frac{\hat{x}}{r_3^2})n_1 - (k_{23}^1 B_1 \frac{\hat{y}_1}{r_1^2} + k_{23}^2 B_2 \frac{\hat{y}_2}{r_2^2} + k_{23}^3 B_3 \frac{\hat{y}_3}{r_3^2})n_2, \quad (A39)$$

$$T_{31} = (k_{21}^1 C_1 \frac{\hat{y}_1}{r_1^2} + k_{21}^2 C_2 \frac{\hat{y}_2}{r_2^2} + k_{21}^3 C_3 \frac{\hat{y}_3}{r_3^2})n_1 + (k_{24}^1 C_1 \frac{\hat{x}}{r_1^2} + k_{24}^2 C_2 \frac{\hat{x}}{r_2^2} + k_{24}^3 C_3 \frac{\hat{x}}{r_3^2})n_2, \quad (A40)$$

$$T_{32} = (k_{24}^1 C_1 \frac{\hat{x}}{r_1^2} + k_{24}^2 C_2 \frac{\hat{x}}{r_2^2} + k_{24}^3 C_3 \frac{\hat{x}}{r_3^2})n_1 + (k_{22}^1 C_1 \frac{\hat{y}_1}{r_1^2} + k_{22}^2 C_2 \frac{\hat{y}_2}{r_2^2} + k_{22}^3 C_3 \frac{\hat{y}_3}{r_3^2})n_2, \quad (A41)$$

$$T_{33} = -(k_{25}^1 C_1 \frac{\hat{x}}{r_1^2} + k_{25}^2 C_2 \frac{\hat{x}}{r_2^2} + k_{25}^3 C_3 \frac{\hat{x}}{r_3^2})n_1 - (k_{23}^1 C_1 \frac{\hat{y}_1}{r_1^2} + k_{23}^2 C_2 \frac{\hat{y}_2}{r_2^2} + k_{23}^3 C_3 \frac{\hat{y}_3}{r_3^2})n_2, \quad (A42)$$

$$U_{11} = A_1 \ln r_1 + A_2 \ln r_2 + A_3 \ln r_3, \quad (A43)$$

$$U_{12} = \alpha_{11} A_1 \arctan \frac{\hat{x}}{\hat{y}_1} + \alpha_{21} A_2 \arctan \frac{\hat{x}}{\hat{y}_2} + \alpha_{31} A_3 \arctan \frac{\hat{x}}{\hat{y}_3}, \quad (A44)$$

$$U_{13} = \alpha_{12} A_1 \arctan \frac{\hat{x}}{\hat{y}_1} + \alpha_{22} A_2 \arctan \frac{\hat{x}}{\hat{y}_2} + \alpha_{32} A_3 \arctan \frac{\hat{x}}{\hat{y}_3}, \quad (A45)$$

$$U_{21} = -B_1 \arctan \frac{\hat{x}}{\hat{y}_1} - B_2 \arctan \frac{\hat{x}}{\hat{y}_2} - B_3 \arctan \frac{\hat{x}}{\hat{y}_3}, \quad (A46)$$

$$U_{22} = \alpha_{11} B_1 \ln r_1 + \alpha_{21} B_2 \ln r_2 + \alpha_{31} B_3 \ln r_3, \quad (A47)$$

$$U_{23} = \alpha_{12} B_1 \ln r_1 + \alpha_{22} B_2 \ln r_2 + \alpha_{32} B_3 \ln r_3, \quad (A48)$$

$$U_{31} = -C_1 \arctan \frac{\hat{x}}{\hat{y}_1} - C_2 \arctan \frac{\hat{x}}{\hat{y}_2} - C_3 \arctan \frac{\hat{x}}{\hat{y}_3}, \quad (A49)$$

$$U_{32} = \alpha_{11} C_1 \ln r_1 + \alpha_{21} C_2 \ln r_2 + \alpha_{31} C_3 \ln r_3, \quad (A50)$$

$$U_{33} = \alpha_{12} C_1 \ln r_1 + \alpha_{22} C_2 \ln r_2 + \alpha_{32} C_3 \ln r_3, \quad (A51)$$

where

$$\hat{x} = x - x_0, \quad \hat{y}_i = s_i(y - y_0), \tag{A52}$$

$$r_i^2 = \hat{x}^2 + \hat{y}_i^2, \tag{A53}$$

and $k_{1j}^i$, $k_{2j}^i$, $a_{i1}$, $a_{i2}$, $A_i$, $B_i$, $C_i$, and $s_i$ $(i = 1, 2, 3, j = 1, 2, ..., 5)$ are the parameters related to the material constants as defined in Refs. [27, 28].